\documentclass[twocolumn,twocolappendix]{aastex701}

\usepackage{color}

\newcommand{\Msun}{M_\sun}
\newcommand{\Mstar}{M_\star}

\begin{document}

\title{CROCODILE-DWARF: Assembly and Kinematics of Field Dwarf Galaxies with GADGET4-OSAKA}

\correspondingauthor{Kazuki Tomaru}
\email{tomaru@astro-osaka.jp}

\author[orcid=0009-0007-4378-406X,gname=Kazuki,sname=Tomaru]{Kazuki Tomaru}
\affiliation{Theoretical Astrophysics, Department of Earth \& Space Science, Graduate School of Science, The University of Osaka, 1-1 Machikaneyama, Toyonaka, Osaka 560-0043, Japan}
\email{tomaru@astro-osaka.jp}

\author[0000-0002-5712-6865,gname=Yuri,sname=Oku]{Yuri Oku}
\affiliation{Center for Cosmology and Computational Astrophysics (C3A), 
Institute for Advanced Study in Physics, Zhejiang University, Hangzhou 310027, People’s Republic of China}
\email{oku@zju.edu.cn}

\author[0000-0003-3467-6079,gname=Daisuke,sname=Toyouchi]{Daisuke Toyouchi}
\affiliation{Theoretical Astrophysics, Department of Earth \& Space Science, Graduate School of Science, The University of Osaka, 1-1 Machikaneyama, Toyonaka, Osaka 560-0043, Japan}
\affiliation{Theoretical Joint Research, Forefront Research Center, Graduate School of Science, The University of Osaka, 1-1 Machikaneyama, Toyonaka, Osaka 560-0043, Japan}
\email{toyouchi@astro-osaka.jp}

\author[0000-0001-7457-8487,gname=Kentaro,sname=Nagamine]{Kentaro Nagamine}
\affiliation{Theoretical Astrophysics, Department of Earth \& Space Science, Graduate School of Science, The University of Osaka, 1-1 Machikaneyama, Toyonaka, Osaka 560-0043, Japan}
\affiliation{Theoretical Joint Research, Forefront Research Center, Graduate School of Science, The University of Osaka, 1-1 Machikaneyama, Toyonaka, Osaka 560-0043, Japan}
\affiliation{Kavli IPMU (WPI), UTIAS, The University of Tokyo, Kashiwa, Chiba 277-8583, Japan}
\affiliation{Department of Physics \& Astronomy, University of Nevada, Las Vegas, 4505 S. Maryland Pkwy, Las Vegas, NV 89154-4002, USA}
\affiliation{Nevada Center for Astrophysics, University of Nevada, Las Vegas, 4505 S. Maryland Pkwy, Las Vegas, NV 89154-4002, USA}
\email{kn@astro-osaka.jp}

\begin{abstract}

We present results from CROCODILE-DWARF, a new suite of cosmological zoom-in hydrodynamic simulations of isolated field dwarf galaxies with halo masses of $\sim10^{10}\,\Msun$ at $z=0$, performed with the \textsc{gadget4-osaka} code. The simulations include detailed modeling of star formation, chemical enrichment, and supernova feedback using the \textsc{CELib} and \textsc{grackle} libraries, achieving baryonic resolutions of $\sim2\times10^3\,\Msun$. Our study focuses on how assembly history governs the structural and kinematic diversity of dwarf galaxies within the $\Lambda$CDM framework. The simulated galaxies reproduce the observed stellar-to-halo mass, mass--metallicity, and size--mass relations for nearby dwarf galaxies, including those of the Local Group, yielding stellar masses of $\sim10^7\,\Msun$. The galaxies display a broad range of rotational support, where gas is generally more rotationally supported than stars.  Differences in morphology and kinematics primarily reflect variations in halo assembly timescales and merger activity. Early-assembling, high-concentration halos form stars efficiently and become gas-poor by $z=0$, while late-assembling, low-concentration halos remain gas-rich due to delayed star formation and rejuvenated gas accretion. We find a trend between rotational support and the cumulative merger mass fraction, providing tentative evidence that dynamical heating induced by mergers plays a role in shaping the kinematic diversity. In some cases, late-time mergers induce the formation of extended gas disks by delivering fresh gas and angular momentum. These results demonstrate that it is assembly history, rather than halo mass alone, that shapes the present-day kinematic and morphological diversity of dwarf galaxies.

\end{abstract}

\keywords{\uat{Dwarf galaxies}{416} --- \uat{Galaxy evolution}{594} --- \uat{Galaxy formation}{595} --- \uat{Galaxy dynamics}{591} --- \uat{Galaxy kinematics}{602} --- \uat{Hydrodynamical simulations}{767}}

\section{Introduction} \label{sec:introduction}

Dwarf galaxies ($\Mstar\lesssim10^9\,\Msun$) serve as important laboratories for probing the distribution and nature of dark matter on both cosmological and galactic scales. These systems exhibit the lowest stellar-to-halo mass ratios ($\Mstar/M_\mathrm{halo}\sim10^{-5}–10^{-3}$), making them particularly sensitive to the effects of dark matter. Furthermore, some dwarf galaxies are satellites of the Milky Way and lie sufficiently close for detailed observations of individual stars and their motions, enabling precise determinations of their mass distributions \citep[e.g.,][]{BullockBoylan-Kolchin2017,Simon2019}.

Since the late 20th century, tensions have arisen between predictions from numerical simulations based on the Lambda cold dark matter ($\Lambda$CDM) model, the standard paradigm of modern cosmology, and observations of dwarf galaxies, sparking a debate. Well-known examples include the ``core--cusp" problem \citep{FloresPrimack1994,Moore1994}, the ``missing satellites" problem \citep{Klypin1999,Moore1999}, and the ``too big to fail" problem \citep{Boylan-Kolchin2011,Boylan-Kolchin2012}. Some of these discrepancies are being partially mitigated within the cold dark matter (CDM) paradigm through the incorporation of baryonic physics in hydrodynamic simulations  \citep{Navarro1996,PontzenGovernato2014,Sawala2016,Wetzel2016,Buck2019,Jung2024}. However, they remain unresolved in full \citep[see, e.g.,][for a review]{Sales2022}. As an alternative, models of dark matter beyond CDM have also been proposed as potential solutions to these small-scale challenges \citep[e.g.,][]{Lovell2012,Vogelsberger2012,Schive2014}.

These problems remain challenging to resolve, largely because dwarf galaxies are highly sensitive to various baryonic processes. Small differences in modeling these processes and their nonlinear interactions can lead to significant uncertainties in theoretical predictions \citep{Munshi2019, Agertz2020}. Moreover, degeneracies between observed properties and the underlying physics further complicate interpretation. Supernova (SN) explosions can readily expel gas from the shallow gravitational potentials of dwarf galaxies \citep[e.g.,][]{DekelSilk1986,Efstathiou2000,Sawala2010,PontzenGovernato2012,Hopkins2018a,Hu2019}, while stellar winds and radiation influence the terminal momentum of superbubbles and the strength of galactic winds even prior to the SN explosions. These feedback processes can enhance SN efficiency by lowering local gas densities, or suppress nearby star formation \citep{Agertz2013,Emerick2018,Kimm2018,Agertz2020,Hopkins2020,Smith2021}. Furthermore, cosmic reionization strips star-forming gas from low-mass dwarf galaxies ($M_\mathrm{halo}\lesssim10^8\,\Msun$) at high redshift, delaying gas cooling and suppressing further gas accretion \citep[e.g.,][]{Rees1986,Efstathiou1992,Shapiro1994,ThoulWeinberg1996,GnedinKaurov2014,Katz2020}. Its influence can persist, strongly inhibiting star formation in halos that remain below the evolving redshift-dependent mass threshold \citep{Gnedin2000,Hoeft2006,Okamoto2008,Benitez-Llambay2020}. More recently, additional processes such as active galactic nucleus feedback, cosmic rays, and magnetic fields have also been recognized as potentially important factors in shaping the formation and evolution of dwarf galaxies \citep[e.g.,][]{Uhlig2012,Chen2016,Pakmor2016,Silk2017,Koudmani2019,Sanati2020,Sanati2024,Hopkins2021,Haidar2022,Martin-Alvarez2023,Arjona-Galvez2024}.

While baryonic processes in dwarf galaxies have been extensively studied theoretically over the past decades, comparatively little attention has been paid to the role of assembly history in shaping their properties. This is likely due to computational cost limitations that make it challenging to generate large samples of dwarf galaxies in cosmological hydrodynamic simulations, which are essential for investigating these processes within a self-consistent framework. However, understanding the role of assembly history will be crucial for rigorous comparisons between theoretical predictions and observations, because assembly history should introduce diversity into dwarf galaxy properties.

Several studies have investigated how the assembly history introduces diversity in dwarf galaxy properties through cosmic reionization/ultraviolet background (UVB) radiation, as well as stellar feedback. This appears to hinge on when halos exceed the halo mass threshold set by cosmic reionization, and their behavior is linked to star formation history \citep[e.g.,][]{Benitez-Llambay2015,Rey2020,Gutcke2022}. In particular, \citet{Fitts2017} conducted zoom-in simulations of dwarf galaxies with a narrow halo mass range of $M_\mathrm{halo}\sim10^{10}\,\Msun$ at $z=0$ and found that the stellar masses strongly correlate with halo properties indicative of formation time, such as the maximum circular velocity and concentration at $z=0$ \citep[see also][]{Dutton2017,Rey2019}.

The morphological and kinematic properties of dwarf galaxies can offer important insights into the underlying physics of galaxy formation, including the role of assembly. They can serve as probes of the burstiness of star formation and feedback strength \citep[e.g.,][]{El-Badry2016,El-Badry2017,El-Badry2018,Yajima2017}. Recently, some studies have begun focusing on the roles of halo or galaxy mergers and interactions, as well as gas accretion from the cosmic web \citep{Dekel2020,Cardona-Barrero2021,Martin2021,Zeng2024}.

Observationally, nearby dwarf galaxies exhibit a wide range of morphologies \citep[e.g.,][]{Mateo1998,McConnachie2012}, with a general trend toward thicker stellar structures as galaxy mass decreases \citep[e.g.,][]{Sanchez-Janssen2010,Roychowdhury2013}. \citet{Simons2015} constructed morphologically unbiased samples of field emission-line galaxies at $z\sim0.2$ and demonstrated that at $\Mstar=10^{9.5}\,\Msun$, galaxies begin to display diverse morphologies, ranging from rotationally supported disks to irregular, asymmetric, or compact shapes, and increasingly fail to form well-defined gas disks.

In more detailed kinematic analyses, \citet{Wheeler2017} studied the stellar rotational support, quantified as the ratio of rotational velocity to velocity dispersion $V_\mathrm{rot,\star}/\sigma_{v,\star}$, in ten isolated Local Group dwarf galaxies ($\Mstar\sim10^5–10^8\,\Msun$), finding that six of the ten are primarily dispersion-supported ($V_\mathrm{rot,\star}/\sigma_{v,\star}<1.0$), while the remainders exhibit only modest rotational support ($V_\mathrm{rot,\star}/\sigma_{v,\star}\lesssim2$), suggesting that dwarf galaxies generally lack cold, well-ordered stellar disks \citep[see also][]{delosReyes2023}. 

On the other hand, \textsc{H\,i} kinematics studies from the  LITTLE THINGS \citep{Hunter2012} show that in some low-mass dwarf galaxies ($\Mstar\sim10^6–10^7\,\Msun$), the neutral gas can exhibit relatively high rotational velocities ($V_\mathrm{rot}\gtrsim30\,\mathrm{km}\,\mathrm{s}^{-1}$) (\citealt{Oh2015,Iorio2017}; see also Section~8.2 in \citealt{El-Badry2018}). Similarly, H$\alpha$ surveys such as SH$\alpha$DE \citep{Barat2020} and EMPRESS \citep{Isobe2023,Xu2024} indicate that some systems show $V_\mathrm{rot}/\sigma_v\sim1$ in ionized gas, despite large velocity dispersions ($\sigma_v\approx15–30\,\mathrm{km}\,\mathrm{s}^{-1}$). These kinematic features are thought to be connected to differences in star formation histories \citep[e.g.,][]{Weisz2011,Weisz2014}, as well as variations in present-day star formation rates, gas content, and metallicity \citep[e.g.,][]{McQuinn2020,McQuinn2021,Nakajima2024arXiv}.

In this study, we perform cosmological zoom-in simulations of dwarf galaxies residing in halos with masses of $M_\mathrm{halo}\sim10^{10}\,\Msun$ at $z=0$. Our work uniquely investigates 
how assembly history drives the diversity in kinematic properties. Following the halo selection strategy of \citet{Wang2015} and \citet{Fitts2017}, we deliberately choose isolated halos with diverse assembly histories. This enables a controlled comparison that isolates the effects of assembly from those of environment and halo mass. Our simulation suite comprises zoom-in simulations performed using the same code as the CROCODILE project \citep{OkuNagamine2024}.

Halos at the $M_\mathrm{halo}\sim10^{10}\,\Msun$ mass scale (and below), corresponding to $\Mstar\lesssim10^7\,\Msun$, lie at the faint end of the baryonic Tully--Fisher relation \citep{McGaugh2000}, a key empirical scaling relation that connects baryonic content and kinematics across galaxy masses. Understanding the behavior of galaxies in this regime provides crucial insights into both baryonic processes and the nature of dark matter \citep[e.g.,][]{McQuinn2022,Sardone2024}.

This paper is organized as follows. In Section~\ref{sec:simulations}, we describe the simulation code and the setup of the initial conditions. Section~\ref{sec:results} presents the main results of our study. In Section~\ref{sec:discussions}, we discuss the physical implications of the findings. Finally, Section~\ref{sec:conclusions} summarizes our conclusions.

\section{Simulations} \label{sec:simulations}

We perform a series of cosmological zoom-in hydrodynamic simulations using the $N$-body smoothed particle hydrodynamics (SPH) code \textsc{gadget4-osaka} \citep{Romano2022a,Romano2022b,OkuNagamine2024}, a successor to \textsc{gadget3-osaka} \citep{Aoyama2017,Shimizu2019,Nagamine2021,Oku2022,Fukushima2023} built upon the \textsc{gadget-4} code \citep{Springel2021}\footnote{\url{https://wwwmpa.mpa-garching.mpg.de/gadget4/}}. For detailed descriptions of the gravitational dynamics and hydrodynamics solvers implemented in the code, we refer the reader to  \citet{Springel2021} and \citet{OkuNagamine2024}.

Radiative gas cooling and heating are modeled using the \textsc{grackle} chemistry and cooling library \citep{Smith2017}\footnote{We use the \textsc{grackle}-3.2.1; see \url{https://grackle.readthedocs.io/en/grackle-3.2.1/}.}. This library calculates cooling and heating rates by solving a nonequilibrium primordial chemical network that includes hydrogen, deuterium, and helium species, along with molecular hydrogen and hydrogen deuteride. For metal-line cooling, \textsc{grackle} employs precomputed tables generated by the photoionization code \textsc{Cloudy} \citep{Ferland2013}, which incorporate photoheating and photoionization effects from the UVB, including self-shielding corrections in high-density gas. We adopt the spatially uniform, redshift-dependent UVB model of \citet{HaardtMadau2012}, included in \textsc{grackle}, and activate it from $z=15$ onward.

\subsection{Star Formation and Supernova Feedback}

Star formation is modeled following a Schmidt-type relation \citep{Schmidt1959}, triggered when the hydrogen number density exceeds $n_\mathrm{H,thres}=1\,\mathrm{cm}^{-3}$ and the gas temperature falls below $T_\mathrm{thres}=10^4\,\mathrm{K}$. The star formation rate density is given by $d\rho_\star/dt=c_\star\rho_\mathrm{gas}/t_\mathrm{ff}$, where $c_\star=0.01$ is the star formation efficiency, $\rho_\mathrm{gas}$ is the local gas density, and $t_\mathrm{ff}={(3\pi/(32G\rho_\mathrm{gas}))}^{1/2}$ is the local freefall time with the gravitational constant $G$. Gas particles that satisfy these criteria are stochastically converted into star particles of equal mass, with the conversion probability set by the above rate. Each star particle represents a simple stellar population, characterized by a single age and metallicity, and assumes the Chabrier initial mass function (IMF; \citealt{Chabrier2003}).

Each star particle distributes feedback energy, mass, and metals to its surroundings from Type II SNe, Type Ia SNe, and asymptotic giant branch (AGB) stars, incorporating appropriate time delays. This is implemented according to the feedback model proposed by \citet{Oku2022}, which utilizes the chemical evolution library \textsc{CELib} \citep{Saitoh2017}. 

SN feedback energy is injected into neighboring gas particles in both momentum and thermal forms, at fixed, logarithmically spaced time intervals. The imparted momentum is computed as a function of the local gas density and metallicity, based on results from three-dimensional hydrodynamic simulations of superbubbles driven by clustered SNe \citep{Oku2022}. The thermal energy is stochastically deposited following a scheme similar to that of \citet{DallaSchaye2012}, heating neighboring gas particles to a target entropy of $K_\mathrm{OF}=10^8\,k_\mathrm{B}\,\mathrm{K}\,\mathrm{cm}^2$ \citep{Hu2019,Keller2020}, where $k_\mathrm{B}$ is the Boltzmann constant. As a parameter tuning measure (see Section~\ref{sec:mzr} and Appendix~\ref{sec:tuning}), we adopt a doubled SN energy budget from the fiducial \textsc{CELib} output, setting the specific SN energy to $\zeta_\mathrm{SN}=2.3\times10^{49}\,\mathrm{erg}\,\Msun^{-1}$. Thereby, both the momentum and thermal energy are increased by a factor of 2 compared with \citet{Oku2022}. This enhancement can be interpreted as a proxy for unresolved early stellar feedback processes preceding SNe, as well as a correction for the limited numerical resolution.

Feedback is distributed to $N_\mathrm{ngb,FB}=120$ neighboring gas particles. The number of feedback events per star particle is set to $n_\mathrm{event,SNII}=2$ for Type II SNe, $n_\mathrm{event,SNIa}=1$ for Type Ia SNe, and $n_\mathrm{event,AGB}=8$ for AGB stars. When computing gas cooling/heating rates and local metallicities for feedback calculations, we employ a smoothed metallicity scheme using the SPH kernel \citep{Okamoto2005,Tornatore2007,Wiersma2009,Shimizu2019}, rather than explicit metal diffusion.

\begin{deluxetable*}{ccccccccc} \label{tab:halos}
\tablecaption{Properties of the Zoom-in Dwarf Galaxies at $z=0$.}
\tablehead{\colhead{Name} & \colhead{$M_{200}$} & \colhead{$R_{200}$} & \colhead{$R_{1/2,\star}$} & \colhead{$M_\mathrm{gas}$} & \colhead{$\Mstar$} & \colhead{$12+\log{(\mathrm{O}/\mathrm{H})}$} & \colhead{$[\mathrm{Fe}/\mathrm{H}]$} & \colhead{$c_{200,\mathrm{DMO}}$} \\ 
\colhead{} & \colhead{$(10^{10}\,\Msun)$} & \colhead{$(\mathrm{kpc})$} & \colhead{$(\mathrm{kpc})$} & \colhead{$(10^8\,\Msun)$} & \colhead{$(10^7\,\Msun)$} & \colhead{} & \colhead{} & \colhead{}} 
\startdata
Halo185 & $1.2$ & $48$ & $1.0$ & $0.34$ & $4.3$ & $8.24$ & $-0.76$ & $14.0$ \\
Halo196 & $1.3$ & $48$ & $1.5$ & $1.6$ & $1.8$ & $7.95$ & $-1.03$ & $6.4$ \\
Halo211 & $1.1$ & $46$ & $0.9$ & $0.11$ & $2.7$ & $8.06$ & $-0.89$ & $14.1$ \\
Halo218 & $1.1$ & $45$ & $0.8$ & $0.11$ & $1.1$ & $7.97$ & $-1.03$ & $8.0$ \\
Halo224 & $0.99$ & $44$ & $1.1$ & $0.74$ & $2.1$ & $8.13$ & $-0.90$ & $9.1$ \\
Halo234 & $1.1$ & $46$ & $1.5$ & $3.0$ & $2.5$ & $7.85$ & $-1.04$ & $10.1$ \\
Halo256 & $0.92$ & $43$ & $1.3$ & $1.0$ & $1.7$ & $7.95$ & $-1.04$ & $11.8$ \\
Halo273 & $0.82$ & $42$ & $1.1$ & $0.55$ & $1.2$ & $8.00$ & $-1.06$ & $9.1$ \\
Halo291 & $0.76$ & $41$ & $0.9$ & $0.18$ & $1.4$ & $7.94$ & $-0.96$ & $14.1$ \\
Halo307 & $0.73$ & $40$ & $0.7$ & $0.41$ & $0.46$ & $7.71$ & $-1.21$ & $10.5$ \\
Halo316 & $0.72$ & $40$ & $1.1$ & $1.0$ & $0.82$ & $7.72$ & $-1.19$ & $9.3$ \\
Halo324 & $0.66$ & $39$ & $0.8$ & $0.072$ & $0.69$ & $7.98$ & $-1.14$ & $12.8$ \\
AGORA-1e10q\tablenotemark{a} & $0.70$ & $39$ & $0.9$ & $0.043$ & $0.66$ & $8.01$ & $-1.16$ & $11.5$ \\
AGORA-1e10v\tablenotemark{a} & $0.76$ & $40$ & $1.2$ & $1.0$ & $0.59$ & $7.68$ & $-1.30$ & $4.9$ \\
\enddata
\tablecomments{The listed properties are as follows: $M_{200}$ is the virial mass, $R_{200}$ is the virial radius, $R_{1/2,\star}$ is the stellar (three-dimensional) half-mass radius, $M_\mathrm{gas}$ is the mass of gas within 3 times the stellar half-mass radius, $\Mstar$ is the mass of stars within 3 times the stellar half-mass radius, $12+\log{(\mathrm{O}/\mathrm{H})}$ is the mean gas-phase metallicity, $[\mathrm{Fe}/\mathrm{H}]$ is the value at the peak of the metallicity distribution function of stars, and $c_{200,\mathrm{DMO}}$ is the concentration of the halos resimulated with dark matter only, defined using the method with the maximum circular velocity of the halo from \citet{Klypin2011} and \citet{Prada2012}.}
\tablenotetext{a}{The original initial conditions are from the AGORA Project \citep{Kim2014}.}
\end{deluxetable*}

\subsection{Initial Conditions}

We generate the initial conditions using the \textsc{Music} code \citep{Hahn2011}\footnote{\url{https://www-n.oca.eu/ohahn/MUSIC/index.html}} with second-order Lagrangian perturbation theory. The adopted cosmological parameters are consistent with WMAP7/9+SNe+BAO: $\Omega_\mathrm{m}=0.272$, $\Omega_\Lambda=0.728$, $\Omega_\mathrm{b}=0.0455$, $\sigma_8=0.807$, $n_\mathrm{s}=0.961$, and $h=0.702$ \citep{Komatsu2011,Hinshaw2013}. 

First, we conduct a preliminary dark-matter-only simulation in a comoving box of ${(10\,h^{-1}\,\mathrm{Mpc})}^3$ using $512^3$ particles from $z=99$ to $z=0$. At $z=0$, isolated halos with virial masses $M_{200}\sim10^{10}\,\Msun$ are identified  with the friends-of-friends group finder \citep{Davis1985} combined with the \textsc{subfind} algorithm \citep{Springel2001}. Here, $M_{200}$ is the mass within the spherical radius $R_{200}$, at which the mean density is 200 times the critical density. The critical density is given by $\rho_\mathrm{crit}=3{H(z)}^2/8\pi G$ with the Hubble parameter $H(z)$. To minimize the impact of past environmental effects, such as interactions with similarly massive or more massive galaxies, we apply strict isolation criteria. Halos are separated from any other friends-of-friends (i.e., central) halo with more than half their masses by a distance greater than 7 times the sum of their respective virial radii $R_{200}$ (i.e., more than approximately $0.5–1\,\mathrm{Mpc}$ away). It allows us to focus on the internal processes, such as feedback and merger history, which govern the evolution and observable properties of dwarf galaxies.

Next, we perform zoom-in hydrodynamic simulations down to $z=0$ with an effective resolution of $2\times2048^3$ particles (i.e., level 11 setup). The refinement regions encompass the Lagrangian regions, where all particles within at least $5R_{200}$ of the target halos at $z=0$ exist. That is to avoid coarse particles contaminating the regions within $2R_{200}$, roughly following \citet{Onorbe2014} and \citet{Hopkins2018b}. 

Additionally, we utilize two initial conditions from the AGORA High-resolution Galaxy Simulations Comparison Project \citep{Kim2014}. Their zoom-in halos (AGORA-1e10q and AGORA-1e10v in Table~\ref{tab:halos}) are selected from a ${(5h^{-1}\,\mathrm{Mpc})}^3$ comoving box with the same cosmological parameters but an initial redshift of $z=100$. We verified that these halos satisfy the same isolation criterion and redefined the refinement regions. 

The masses of dark matter and initial gas particles are $m_\mathrm{DM}=1.04\times10^4\,\Msun$ and $m_\mathrm{gas,IC}=2.09\times10^3\,\Msun$. The gravitational softening lengths of refined dark matter and baryon particles remain constant in comoving coordinates until they reach their physical values of $104\,\mathrm{pc}$ and $26\,\mathrm{pc}$ at $z=2.3$ in the Plummer-equivalent values, respectively, after which they remain constant in physical coordinates down to $z=0$. We set the gravitational softening length of dark matter particles according to $\approx0.005(N_{200,\mathrm{DM}}/10^5)^{-1/3}R_{200}\approx0.015l_\mathrm{DM}$, as proposed by \citet{vandenBoschOgiya2018} and \citet{Ludlow2019a,Ludlow2020}, where $N_{200,\mathrm{DM}}$ is the number of dark matter particles within $R_{200}$ and $l_\mathrm{DM}=L_\mathrm{box}/N_{\mathrm{eff,DM}}$ is the mean interparticle distance of dark matter particles with the box side length $L_\mathrm{box}$ and the effective number of dark matter particles $N_{\mathrm{eff,DM}}$. No minimum gas smoothing length is imposed. The initial metallicity floor is $10^{-4}\,Z_\sun$, where $Z_\sun=0.0134$ is the solar metallicity \citep{Asplund2009}. The merger trees are constructed using \textsc{gadget-4} \citep{Springel2021,Springel2022}. The output of the simulation is stored approximately every $50\,\mathrm{Myr}$. Our simulations are summarized in Table~\ref{tab:halos}.

\section{Results} \label{sec:results}

\subsection{Stellar-to-Halo Mass Relation}

\begin{figure}[t]
\centering
\plotone{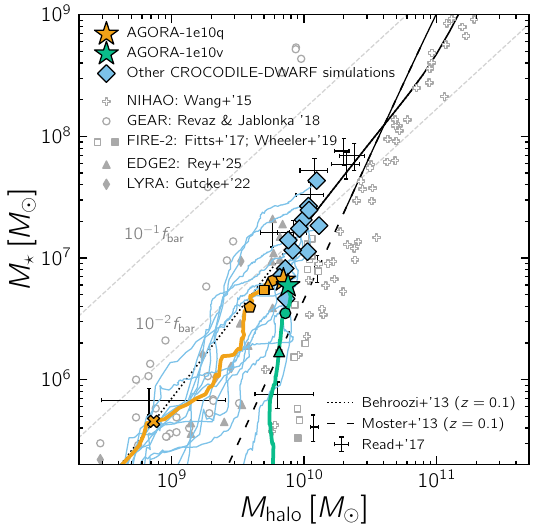}
\caption{Stellar-to-halo mass relation at $z=0$, with evolutionary tracks for dwarf galaxies in our simulations. We highlight two sample galaxies, AGORA-1e10q and AGORA-1e10v, and show their relations at redshifts $z=0.25$, $0.5$, $1.0$, $2.0$, and $5.0$. Our results are compared with empirical relations derived from abundance matching and their extrapolations by \citet{Behroozi2013} and \citet{Moster2013} at $z=0.1$, as well as with the relation inferred from \textsc{H\,i} rotation curve observations of field dwarf galaxies \citep{Read2017}. For reference, we also show results from other simulations in the literature, along with lines corresponding to $10\,\%$ and $1\,\%$ of the baryon fraction of the universe $f_\mathrm{bar}$.}
\label{fig:shmr}
\end{figure}

Figure~\ref{fig:shmr} presents the stellar-to-halo mass relation at $z=0$ for our simulations, indicated by blue diamonds. We define the stellar mass of a galaxy as the total mass of star particles located within 3 times the stellar half-mass radius from its center. The stellar half-mass radius is defined as the radius of the sphere that encloses half of the total mass of gravitationally bound star particles. The center of the galaxy is identified as the position of the particle with the lowest gravitational potential among all gravitationally bound particles.

We compare our results with empirical stellar-to-halo mass relations from abundance matching by \citet{Behroozi2013} and \citet{Moster2013} at $z=0.1$, shown as solid black lines, with their extrapolations indicated by dotted and dashed lines, respectively. We also include the relation derived from \textsc{H\,i} rotation curve observations of field dwarf galaxies by \citet{Read2017}. Furthermore, we compare against recent simulations: NIHAO \citep[][pluses]{Wang2015}, GEAR \citep[][circles]{RevazJablonka2018}, FIRE-2 \citep[][squares]{Fitts2017,Wheeler2019}, EDGE2 \citep[][triangles]{Rey2025}, and LYRA \citep[][thin diamonds]{Gutcke2022}, all plotted as gray symbols. We note that the definition of halo mass differs among these works: \citet{Behroozi2013} and FIRE-2 \citep{Fitts2017,Wheeler2019} adopt the virial definition of \citet{BryanNorman1998}, whereas the other studies use $M_{200}$. We plot all data using their original halo mass definitions without corrections to maintain transparency, as applying corrections would require some assumptions, such as about halo concentration, not available for all studies.

Our results at $z=0$ are broadly consistent with both abundance matching results and observational constraints. While they lie slightly above the relation of \citet{Moster2013}, the discrepancy is negligible given the substantial uncertainties in the stellar-to-halo mass relation for low-mass dwarf galaxies, which remain challenging to constrain observationally. On the theoretical side, recent simulations predict stellar masses that span more than two orders of magnitude for halos with  $M_\mathrm{halo}\lesssim10^{10}\,\Msun$ \citep{Sales2022}, and our results fall well within this range.

Additionally, Figure~\ref{fig:shmr} presents the evolutionary tracks of our sample and highlights two representative galaxies, AGORA-1e10q and AGORA-1e10v, and shows their relations at redshifts $z=0.25$, $0.5$, $1.0$, $2.0$, and $5.0$, represented by small circles, triangles, squares, pentagons, and crosses, respectively. AGORA-1e10q follows a typical evolutionary track within our sample, evolving roughly along the stellar-to-halo mass relation at $z=0$. In contrast, AGORA-1e10v accretes more of its mass at later times, experiences delayed star formation, and exhibits a remarkably steep evolutionary track.

\subsection{Mass--Metallicity Relations} \label{sec:mzr}

\begin{figure}[t]
\centering
\plotone{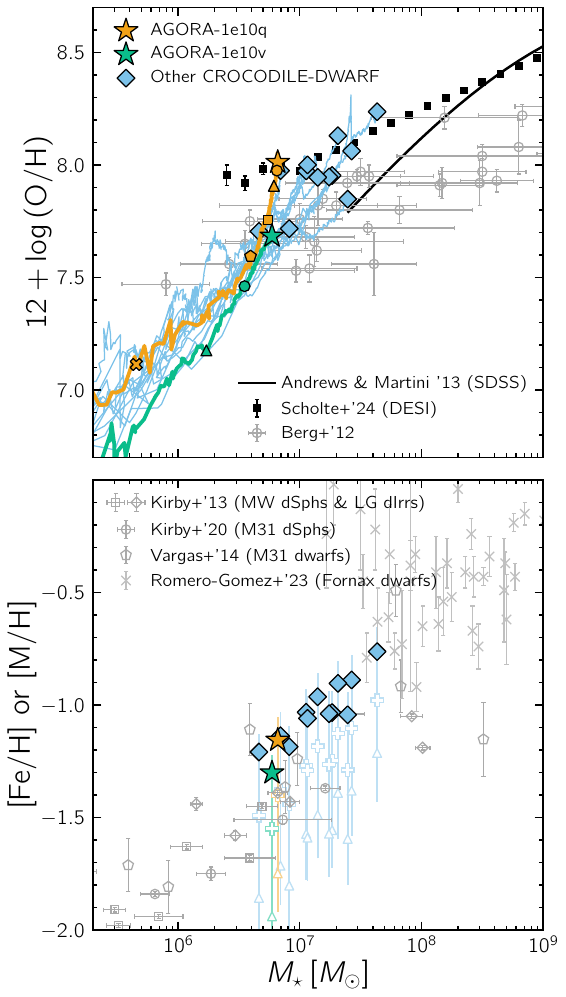}
\caption{Mass--metallicity relations at $z=0$ for dwarf galaxies in our simulations. The top panel shows the relation for gas, while the bottom panel shows that for stars. In the bottom panel, light-colored small triangles, small pluses, and error bars represent the mean, median, and 25th--75th percentile ranges of stellar metallicities of star particles, respectively. Observational data for gas are taken from the SDSS sample \citep{AndrewsMartini2013}, the DESI sample \citep{Scholte2024}, and individual nearby galaxies \citep{Berg2012}, while those for stars are from Local Group dwarf galaxies \citep{Kirby2013,Kirby2020,Vargas2014} and Fornax dwarf galaxies \citep{Romero-Gomez2023}.}
\label{fig:mzr}
\end{figure}

Figure~\ref{fig:mzr} compares the gas-phase and stellar metallicities at $z=0$ for our simulations with observational data. The symbols for the simulations are the same as in Figure~\ref{fig:shmr}. The top panel shows the gas-phase metallicities, while the bottom panel shows the stellar metallicities. For the gas-phase metallicity, as in Figure~\ref{fig:shmr}, the evolution tracks are shown.

We define galaxy metallicities in our simulations as follows. The gas-phase metallicity is defined as the mass-weighted mean oxygen-to-hydrogen number ratio of gas particles within 3 times the stellar half-mass radius from the galactic center. The stellar metallicity is defined as the value at the peak of the mass-weighted distribution of iron abundances for star particles in the same region, as described by \citet{Sanati2023}. The peak is determined by fitting the distribution with the functional form predicted by a simple galactic chemical evolution model \citep{Schmidt1963,SearleSargent1972}. Star particles with $[\mathrm{Fe}/\mathrm{H}]<-3$ are excluded from the peak measurement. For reference, we also show the mean, median, and 25th--75th percentile ranges of iron abundances of star particles as light-colored small triangles, small pluses, and error bars, respectively.

For gas-phase metallicities, we compare with the relation for star-forming galaxies from the Sloan Digital Sky Survey (SDSS; \citealt{AndrewsMartini2013}, black solid line), measurements from the Dark Energy Spectroscopic Instrument (DESI; \citealt{Scholte2024}, black squares with error bars), and data for individual nearby galaxies \citep[][gray circles]{Berg2012}. For stellar metallicities, we use data for Local Group dwarf galaxies (\citealt{Kirby2013}, squares and diamonds; \citealt{Kirby2020}, circles; \citealt{Vargas2014}, pentagons) and for the Fornax dwarf galaxies (gray crosses; \citealt{Romero-Gomez2023}). For the \cite{Vargas2014} sample, we take the stellar masses from \cite{McConnachie2012}.

We do not define galaxy stellar metallicity as the mean iron abundance of star particles \citep[see Equations~(3) and (4) in][]{Escala2018}, a method motivated by observational measurements of Local Group dwarf galaxies. This avoids underestimating metallicity due to the inclusion of excessively metal-poor stars in the simulations, as discussed in \citet{Sanati2023}. \citet{Escala2018} demonstrated that simulations without an explicit subgrid model of metal diffusion exhibit broader metallicity distribution functions than those with it, with tails composed of excessively low-metallicity star particles. For technical reasons (e.g., substantial computational costs), we did not turn on the metal diffusion model for the simulations presented in this paper. As a result, they exhibit similar low-metallicity tails, leading to a systematic offset of approximately $0.5–0.7\,\mathrm{dex}$ between the peak and mean definitions of galaxy stellar metallicity (see diamonds and triangles in Figure~\ref{fig:mzr}). Since the mean is computed by averaging the logarithms of the metallicities, it is susceptible to the presence of low-metallicity star particles.

Recently, \citet{Agertz2020} argued that the stellar mass-metallicity relation is an important indicator for constraining feedback models, performing zoom-in simulations of an isolated dwarf galaxy with a halo mass $M_\mathrm{halo}\sim10^9\,\Msun$ at $z=0$. In our simulations, we have doubled the energy of SN feedback injected into galaxies as part of our parameter tuning process (see Appendix~\ref{sec:tuning}). Notably, the standard energy before enhancement overestimated the metallicity of galaxies, consistent with the finding of \citet{Agertz2020}. Our results overlap well with the observations, indicating that the feedback in our simulations is energetically sufficient and does not overpredict the metallicities. The evolutionary tracks in gas-phase metallicity for our simulated galaxies follow the observed relation at $z=0$.

\subsection{Galaxy Sizes}

\begin{figure}[t]
\centering
\plotone{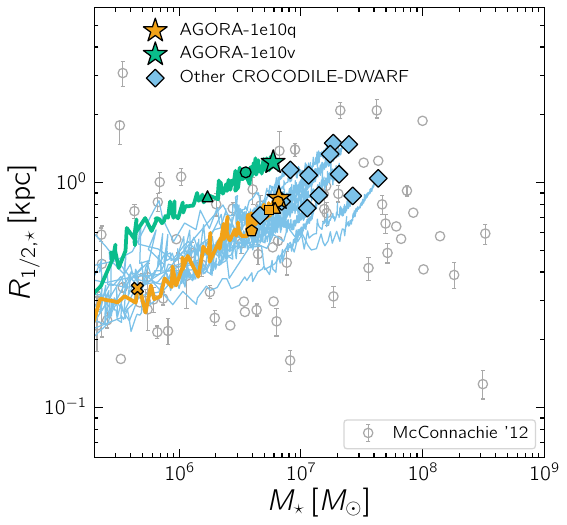}
\caption{Stellar half-mass radius as a function of stellar mass for dwarf galaxies in our simulations. The observational data are from the Local Group and nearby dwarf galaxies \citep{McConnachie2012}.}
\label{fig:size}
\end{figure}

Figure~\ref{fig:size} shows the stellar half-mass radius as a function of stellar mass for our simulations at $z=0$. For comparison, we include observational data from nearby dwarf galaxies \citep{McConnachie2012}\footnote{We use the updated version on 2021 January 19; see \url{https://www.cadc-ccda.hia-iha.nrc-cnrc.gc.ca/en/community/nearby/}.}. To compare with the stellar half-mass radius from our simulations, we convert the observed half-light radius measured along the semimajor axis by multiplying it by $4(1-e)^{1/2}/3$, where $e$ is the ellipticity, following \citet{Sales2022}.

Our simulations exhibit sizes of $R_{1/2,\star}=0.7–1.5\,\mathrm{kpc}$ in the stellar mass range $\Mstar=10^7–10^8\,\Msun$. While these sizes fall within the observed range, our simulations tend to lie in the upper portion of the observed distribution. The observed dwarf galaxies exhibit a significant scatter in size at a fixed mass, with some galaxies being substantially more compact than our simulations predict. This trend may arise from the choice of numerical parameters. In particular, the dark matter gravitational softening length of $104\,\mathrm{pc}$ limits our ability to fully resolve the central regions of halos, potentially leading to underestimated central densities. Furthermore, our adopted star formation threshold density $n_\mathrm{H,thres}=1\,\mathrm{cm}^{-3}$ may be low, allowing stars to form over more extended regions.

Most of our samples, including AGORA-1e10q, follow a broadly similar evolutionary track. In contrast, AGORA-1e10v exhibits a notably larger size throughout its evolution. This difference is likely due to AGORA-1e10v acquiring most of its mass through late, violent mergers, undergoing significant dynamical heating, and possessing a shallow gravitational potential due to its low halo concentration.

\subsection{Kinematics of Gas and Stars}

\begin{figure}[t]
\centering
\plotone{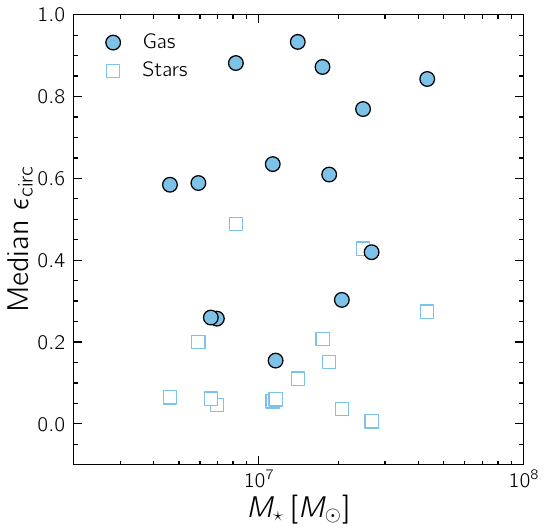}
\caption{Median orbital circularity parameter of gas (circle) and stars (square) at $z=0$ as a function of stellar mass for dwarf galaxies in our simulations.}
\label{fig:circ}
\end{figure}

We utilize the distribution of the ``orbital circularity'' parameter \citep{Abadi2003} to characterize the kinematics and morphology of galaxies for our simulations. The orbital circularity $\epsilon_\mathrm{circ}$ for each gas and star particle $i$ is defined as
\begin{equation}
    \epsilon_{\mathrm{circ},i}=\frac{j_{z,i}}{j_\mathrm{circ}(E_i)},
\end{equation}
where $j_{z}$ is the component of a particle's specific angular momentum aligned with the galaxy's angular momentum vector, and $j_\mathrm{circ}(E)$ is the specific angular momentum of a particle in a circular orbit with the same mechanical energy $E$. We define the galaxy's angular momentum vector separately for gas and stars as the vector sum of the angular momenta of particles within 3 times the stellar half-mass radius from the galactic center. When calculating the mechanical energy of particles, we spherically average the gravitational potential as in \citet{El-Badry2018}. In Figure~\ref{fig:circ}, we present the median orbital circularity parameter for our simulations at $z=0$ as a function of stellar mass. Filled circles represent gas and open squares represent stars.

Our simulations reveal trends in the median orbital circularity parameters for gas and stellar components. The gas component consistently exhibits higher $\epsilon_{\mathrm{circ}}$ than the stellar component, reflecting more rotationally supported disk-like structures, while stars maintain more spheroidal, dispersion-dominated distributions. Furthermore, our sample shows almost no dependence of the orbital circularity parameter on stellar mass. The scatter in $\epsilon_{\mathrm{circ}}$ likely reflects the diversity of assembly histories among our simulations, as we discuss in the following section.

The qualitative trend of higher gas circularity compared to stars is consistent with \citet{El-Badry2018}, although our median $\epsilon_\mathrm{circ,gas}\approx0.2–0.9$ is systematically higher than those reported in that work ($\sim0.1–0.5$ for a similar halo mass range). We discuss possible origins of this difference in Section~\ref{sec:gasdisk}.

Unlike dissipative gas, stellar kinematics are governed by collisionless interactions. This difference can mean that stars preserve the imprint of various dynamical processes over cosmic time, resulting in lower circularity relative to the galaxy's current angular momentum vector. Specifically, stars undergo dynamical heating over time and migrate to outer radii through feedback processes \citep{Stinson2009,Maxwell2012,El-Badry2016,El-Badry2017}. This heating occurs due to oscillations in the gravitational potential caused by gas inflows and outflows \citep{PontzenGovernato2012}. Furthermore, the orientation of a galaxy's angular momentum vector can change over time as gas is accreted and realigns with the new infall \citep{Zeng2024}.

Additionally, several numerical effects can influence the circularity distribution. The stochastic thermal energy injection in our SN feedback model produces episodic feedback events that induce fluctuations in the gravitational potential, contributing to dynamical heating. Two-body relaxation can also make the stellar velocity distribution more isotropic \citep{Power2003,Binney2008}. Following \citet{Power2003}, the convergence radius in our simulations is approximately $0.2–0.4\,\mathrm{kpc}$, which is smaller than the stellar half-mass radius, suggesting that two-body relaxation should not significantly affect the bulk stellar kinematics. However, another source of artificial heating arises from energy transfer in two-body scattering between star and dark matter particles that differ in mass \citep[e.g.,][]{Ludlow2019b}. This spurious energy transfer can artificially inflate star particles and produce systems that are erroneously dispersion-dominated. To minimize this effect, we adopt the dark matter gravitational softening length recommended by \citet{Ludlow2019a,Ludlow2020}, which should help mitigate artificial heating from mass segregation in our simulations.

\subsection{Impact of Assembly on Galaxy Masses}

\begin{figure*}[t]
\centering
\includegraphics[width=1.0\textwidth]{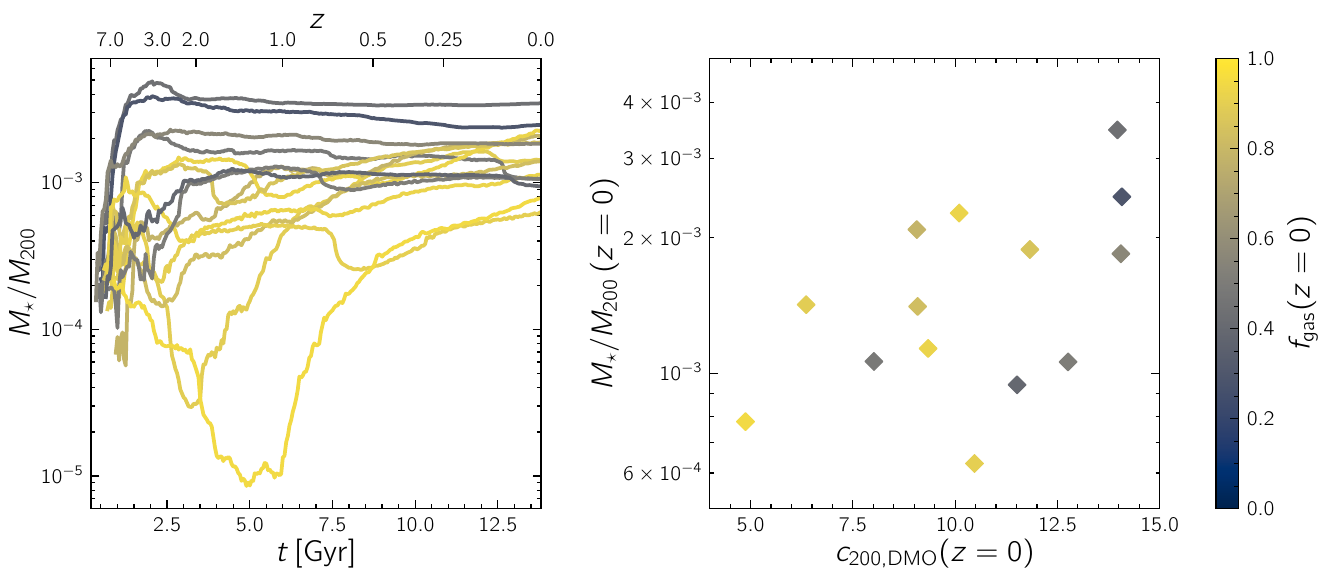}
\caption{\textit{Left panel:} evolution of the stellar-to-halo mass ratio over time for dwarf galaxies in our simulations. \textit{Right panel:} relation between the halo concentration in the dark-matter-only resimulations and the stellar-to-halo mass ratio at $z=0$. The lines in the left panel and symbols in the right panel are color-coded by the gas fraction of galaxies $f_\mathrm{gas}=M_\mathrm{gas}/(M_\mathrm{gas}+\Mstar)$ at $z=0$.}
\label{fig:assembly-mass}
\end{figure*}

Figure~\ref{fig:assembly-mass} shows the evolution of the stellar-to-halo mass ratio over time for dwarf galaxies in our simulations (left panel) and the relation between the halo concentration in the dark-matter-only resimulations and the stellar-to-halo mass ratio at $z=0$ (right panel). The lines in the left panel and the symbols in the right panel are color-coded by the gas fraction of galaxies, defined as $f_\mathrm{gas}=M_\mathrm{gas}/(M_\mathrm{gas}+M_\star)$, at $z=0$. We derive the halo concentration $c_{200,\mathrm{DMO}}$ in the dark matter-only resimulations using the method from \citet{Klypin2011} and \citet{Prada2012}. We specifically solve
\begin{equation}
    \frac{V_\mathrm{max}}{V_{200}}={\left(\frac{0.216\,c_{200}}{f(c_{200})}\right)}^{1/2}
\end{equation}
where $V_\mathrm{max}$ is the maximum circular velocity of the halo, $V_{200}=(GM_{200}/R_{200})^{1/2}$, and $f(c)=\ln{(1+c)}-c/(1+c)$.

The evolutionary tracks of the stellar-to-halo mass ratio reveal diversity among our dwarf galaxies, reflecting their varied assembly histories. Halo concentration serves as a good indicator of the halo's assembly history \citep[e.g.,][]{Wechsler2002}. Halos with the earliest assembly, characterized by the highest concentrations of $c_{200,\mathrm{DMO}}>12$, accumulate most of their mass by $z\sim2–3$. These systems reach stellar-to-halo mass ratios ($\Mstar/M_{200}>10^{-3}$) early in their evolution and exhibit little subsequent growth. As a result, they become gas-poor ($f_\mathrm{gas}<0.5$) by $z=0$, owing to the lack of fresh gas supply from the cosmic web at late times ($z<2$). In contrast, halos with later assembly and lower concentrations, especially $c_{200,\mathrm{DMO}}<7$, follow a different evolutionary path. These systems initially experience quenching of star formation due to the combined effects of reionization and SN feedback at high redshift, when their shallow potential wells cannot retain gas against heating and outflows, also limiting early gas accretion \citep{Onorbe2015}. However, continued mass growth enables renewed gas accretion at $z<3$, allowing them to reignite star formation and maintain substantial gas reservoirs ($f_\mathrm{gas}>0.8$) at $z=0$. Consequently, they end up with lower stellar-to-halo mass ratios ($\Mstar/M_{200}\lesssim10^{-3}$) \citep{Fitts2017}.

The correlation between halo concentration and stellar-to-halo mass ratio thus reflects the two distinct evolutionary pathways. High-concentration halos that assembled early never undergo complete quenching but steadily deplete their gas through continuous star formation, leaving them gas-poor by $z=0$. Low-concentration halos, on the other hand, undergo temporary quenching followed by rejuvenation, as their extended assembly histories permit late-time gas accretion that replenishes their interstellar medium.

For halos with intermediate concentrations ($c_{200,\mathrm{DMO}}\sim10$), the scatter in gas fraction arises from the merger histories. As shown in the left panel of Figure~\ref{fig:assembly-mass}, systems that experience interactions and subsequent mergers can exhibit dips in their evolutionary tracks, acquiring additional gas at $z<3$ and retaining it until $z=0$. However, not all dips correspond to gas-rich mergers. Some are associated with gas-poor mergers that do not significantly increase the gas content.

\subsection{Impact of Assembly on the Kinematics}

\begin{figure}[t]
\centering
\plotone{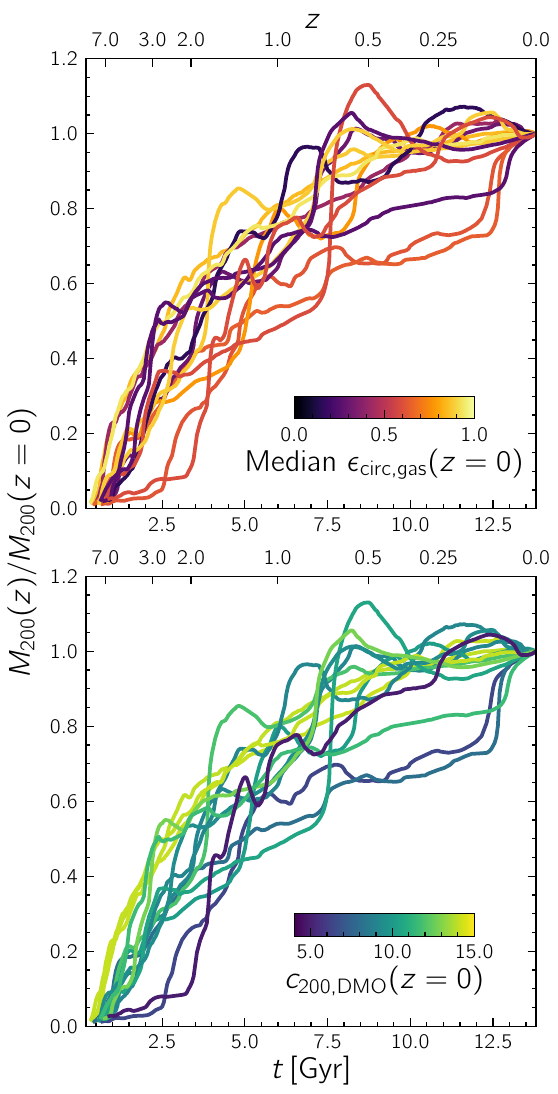}
\caption{Evolution of the virial mass for dwarf galaxies in our simulations, normalized to its value at $z=0$. The lines are color-coded by the median orbital circularity parameter of gas at $z=0$ (upper panel) and the halo concentration at $z=0$ (lower panel).}
\label{fig:massgrowth}
\end{figure}

Figure~\ref{fig:massgrowth} shows the evolution of the virial mass over time for our simulations, normalized to its value at $z=0$. In the upper panel, the lines are color-coded by the median orbital circularity of gas, while in the lower panel, they are color-coded by the halo concentration at $z=0$.

Figure~\ref{fig:massgrowth} indicates no clear connection between the mass growth history and the gas kinematics. While no simple one-to-one correspondence is apparent, the diversity of mass growth histories reveals some suggestive cases. Among the yellow lines (median $\epsilon_\mathrm{circ,gas}>0.8$), two show smooth, monotonic mass growth histories, which correspond to galaxies residing in the highest-concentration halos ($c_{200,\mathrm{DMO}}\approx14$) in our simulations. The other two yellow lines exhibit prominent bumps, corresponding to major merger events that likely triggered the formation of gas disks, as discussed in Section~\ref{sec:gasdisk}. A smooth purple line (median $\epsilon_\mathrm{circ,gas}\approx0.4$; Halo211) is an outlier that experiences a late, close flyby of an infalling satellite halo (dark matter mass ratio $\mu\approx0.04$, $z=0.1–0.2$; $\mu$ is defined in the following paragraph in detail), which may have contributed to its low gas circularity. Despite its high halo concentration ($c_{200,\mathrm{DMO}}=14.1$) and monotonic mass growth history, Halo211 contains little rotationally supported gas, unlike the other halos with similarly high concentrations. Interestingly, some red lines, indicating moderate gas circularity ($\epsilon_\mathrm{circ,gas}\approx0.6$), correspond to the lowest halo concentrations ($c\lesssim8$) in our simulations, likely reflecting recent mass accretion.

We next quantify the cumulative impact of mergers by calculating the fraction of mass accreted through mergers, defined as 
\begin{equation}
    f_\mathrm{acc}=\frac{\sum_i{M_{\mathrm{acc,peak},i}}}{M_\mathrm{main}}
\end{equation}
where $M_{\mathrm{acc,peak},i}$ is the peak dark matter mass of accreted halo $i$ with a mass ratio $\mu>0.01$, and $M_\mathrm{main}$ is the dark matter mass of the main halo. Both masses refer to the subhalo mass identified with the \textsc{subfind} algorithm. $\mu$ is defined as the ratio of the dark matter mass of the accreted halo to that of the main halo at the time when the accreted halo reaches its peak dark matter mass. This metric effectively captures the contribution of both major and minor mergers to the final halo mass during the period when most gases and stellar mass form. To ensure that accreted halos are resolved with at least 100 dark matter particles, we only consider halos that reach their peak mass at $z<7$, when the main halo masses exceed $10^8\,\Msun$.

\begin{figure}[t]
\centering
\plotone{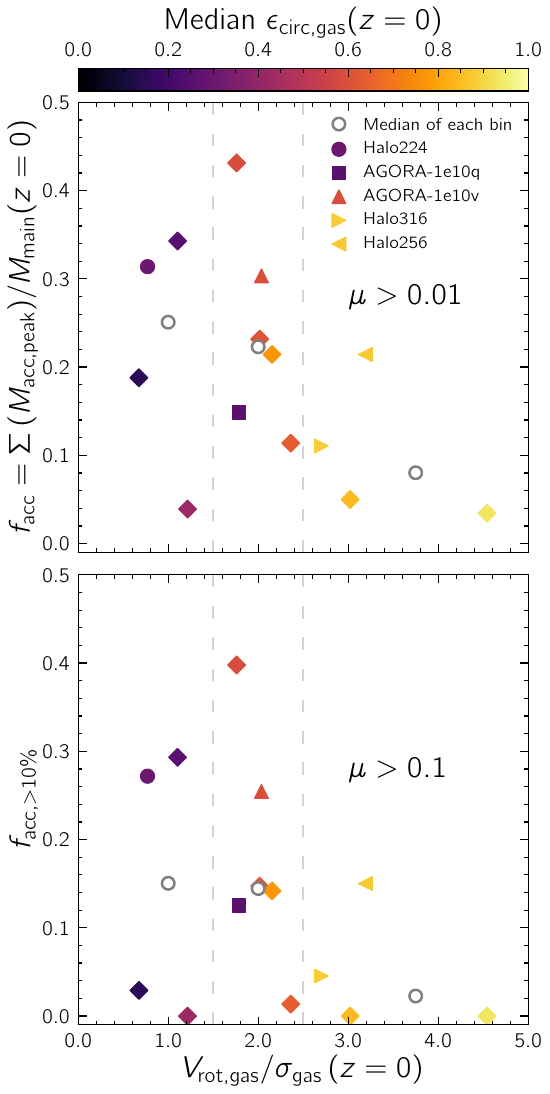}
\caption{Relation between the gas rotational support at $z=0$, and the fraction of mass accreted through mergers $f_\mathrm{acc}$ (upper panel, $\mu>0.01$) and $f_{\mathrm{acc},>10\%}$ (lower panel, $\mu>0.1$) at $z=0$, for dwarf galaxies in our simulations. Open gray circles indicate the median values in $x$-axis bins, while vertical gray dashed lines mark the bin boundaries. Five dwarf galaxies (AGORA-1e10q, AGORA-1e10v, Halo224, Halo316, and Halo256) are highlighted with distinct symbols (circle, square, upward triangle, left-pointing triangle, and downward triangle, respectively).}
\label{fig:rot-merger}
\end{figure}

Figure~\ref{fig:rot-merger} shows the relation between gas rotational support, quantified by the ratio of the gas rotational velocity to the gas velocity dispersion $V_\mathrm{rot,gas}/\sigma_\mathrm{gas}$ at $z=0$, and the merger mass contribution fraction at $z=0$. When calculating $V_\mathrm{rot,gas}/\sigma_\mathrm{gas}$, we adopt cylindrical coordinates aligned with the galaxy's gas angular momentum vector. In each $0.3\,\mathrm{kpc}$ cylindrical bin $j$, we compute the mass-weighted mean azimuthal velocity $V_{\phi,j}$ and the one-dimensional equivalent velocity dispersion $\sigma_{\mathrm{gas},j}=(({\sigma_{\mathrm{gas},r,j}}^2+{\sigma_{\mathrm{gas},\phi,j}}^2+{\sigma_{\mathrm{gas},z,j}}^2)/3)^{1/2}$. We then define $V_\mathrm{rot,gas}$ as the peak value of $V_{\phi,j}$ and $\sigma_\mathrm{gas}$ as the median of $\sigma_{\mathrm{gas},j}$ across all bins. The measurement region extends to 3 times the stellar half-mass radius in the radial direction and one stellar half-mass radius above and below the midplane in the vertical direction.

The upper panel of Figure~\ref{fig:rot-merger} exhibits a trend between $V_\mathrm{rot,gas}/\sigma_\mathrm{gas}$ and $f_\mathrm{acc}$, suggesting that dynamical heating induced by mergers may shape the gas kinematics of dwarf galaxies. Rotationally supported galaxies tend to have lower merger fractions, though with considerable scatter. We find a Spearman correlation coefficient of $-0.5$ ($p=0.06$). We also show $f_{\mathrm{acc},>10\%}$ computed with a more conservative threshold of $\mu>0.1$ in the lower panel. The trend persists but weakens slightly, with a Spearman correlation coefficient of $-0.4$ ($p=0.17$). The stronger correlation obtained for $\mu>0.01$ than for $\mu>0.1$ suggests that ``marginal'' mergers with $\mu=0.01–0.1$ contribute to shaping the kinematics of some galaxies. Nevertheless, the large scatter indicates that, in addition to merger activity, other processes, such as merger timing, interactions with accreting satellites, gas accretion, star formation history, and halo shape, may also play important roles \citep{DowningOman2023}. Because our sample consists of only 14 simulations, a larger sample will be required to improve statistical power and robustly establish the connection between merger history and gas kinematics in dwarf galaxies.

\begin{figure*}[t]
\centering
\plotone{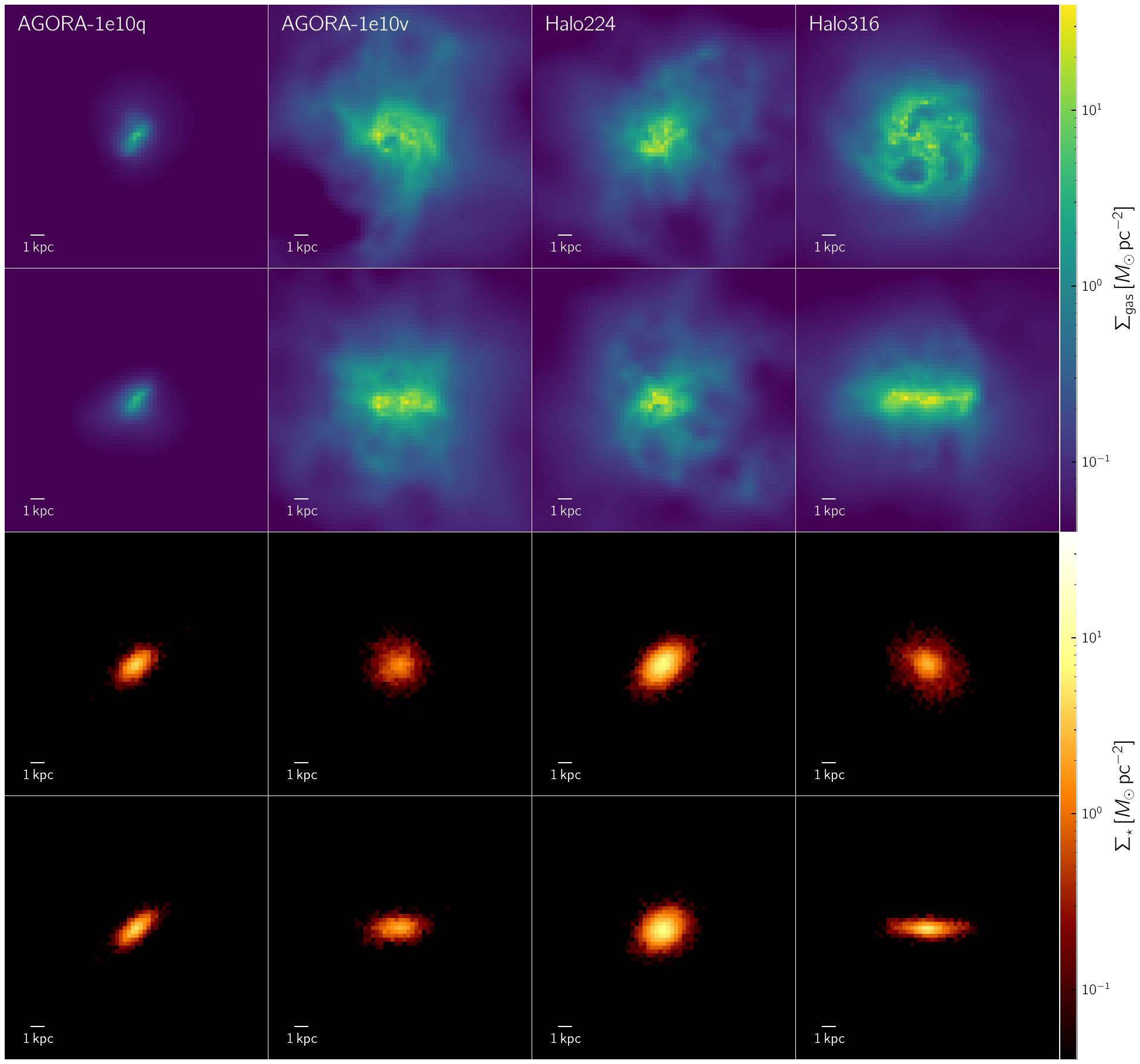}
\caption{Surface density maps for four dwarf galaxies in our simulations (AGORA-1e10q, AGORA-1e10v, Halo224, and Halo316), showing the gas (top two rows) and stellar (bottom two rows) components. For each component, the upper row shows face-on views (aligned with the galaxy's gas angular momentum vector) and the lower row shows edge-on views. Each panel represents the projected density within a ${(20\,\mathrm{kpc})}^3$ volume.}
\label{fig:maps}
\end{figure*}

In Figure~\ref{fig:maps}, to highlight the morphological diversity arising from different assembly histories, we show surface density maps for four dwarf galaxies in our simulations: AGORA-1e10q, AGORA-1e10v, Halo224, and Halo316. These galaxies span a range of rotational support parameters and cumulative merger mass fractions, as highlighted in Figure~\ref{fig:rot-merger}. The maps show the projected densities of gases and stars within a ${(20\,\mathrm{kpc})}^3$ volume, displayed as both face-on views aligned with the galaxy's gas angular momentum vector and edge-on views for each component. AGORA-1e10q represents a gas-poor galaxy in our sample, relatively quiescent but with some mergers ($f_\mathrm{acc}\approx0.15$) and lacking orderly kinematics. AGORA-1e10v and Halo224, which experienced substantial merger activity ($f_\mathrm{acc}\approx0.3$), exhibit moderately disturbed and highly disturbed kinematics, respectively ($V_\mathrm{rot,gas}/\sigma_\mathrm{gas}\lesssim2$). In contrast, Halo316, which has rotational support ($V_\mathrm{rot,gas}/\sigma_\mathrm{gas}\approx3$) and minimal merger activity ($f_\mathrm{acc}\approx0.1$), displays a well-defined disk structure in both gas and stellar components, with the gas forming an extended thin disk clearly visible in the edge-on view.

\section{Discussions} \label{sec:discussions}

\subsection{Frequency and Mass Ratio of Mergers}

The trend between gas rotational support and cumulative merger mass fraction shown in Figure~\ref{fig:rot-merger} suggests that mergers play an important role in shaping the kinematics of dwarf galaxies, though the correlation is not statistically significant given our sample size. Systems that undergo substantial merger activity tend to display dispersion-dominated kinematics, whereas those with relatively quiet merger histories tend to preserve rotationally supported structures.

\citet{Dekel2020} discussed the survival of gas disks analytically using cosmological hydrodynamical simulations based on halo and galaxy major merger rates, and argued that disk survival is difficult in galaxies below a threshold halo mass of $\sim2\times10^{11}\,\Msun$. In mergers, when the orbital angular momentum vector of the accreting object does not align with the disk's angular momentum vector, the disk is not expected to survive. Under the assumption that galaxies are located at nodes of the cosmic web and the accretion directions of objects correlate with directional changes in cosmic web streams, they predicted that for low-mass halos below this threshold mass, disks cannot survive because major mergers occur on timescales shorter than the disk rotation timescale.

Our simulations are qualitatively consistent with their argument. The halo mass is $M_{200}\sim10^{10}\,\Msun$, which is below their mass threshold, and 
10 out of our 14 samples exhibit dispersion-supported gas kinematics ($V_\mathrm{rot,gas}/\sigma_\mathrm{gas}\lesssim2$) as a result of frequent mergers.

\begin{figure}[t]
\centering
\plotone{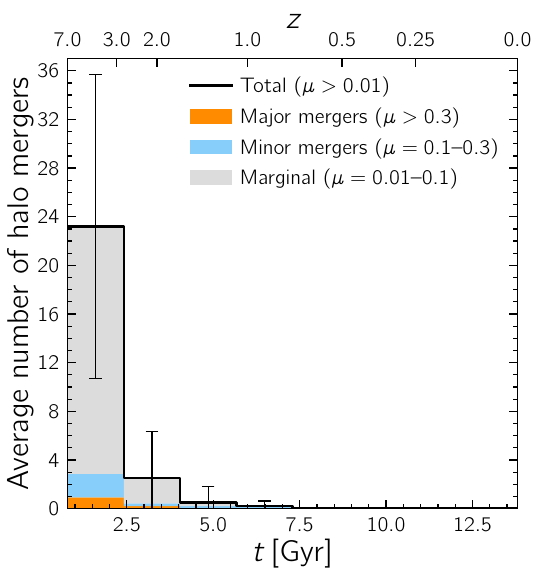}
\caption{Average number of halo mergers ($z<7$) for dwarf galaxies in our simulations. Time indicates when accreted halos reach peak mass. The black-outlined histogram shows mergers with mass ratios $\mu > 0.01$, with the $1\sigma$ range indicated by error bars. Colors indicate the fraction of major mergers ($\mu>0.3$, orange), of minor mergers ($\mu=0.1–0.3$, light blue), and of ``marginal" mergers ($\mu=0.01–0.1$, gray).}
\label{fig:mergers}
\end{figure}

However, major mergers are rare events in dwarf galaxies. Figure~\ref{fig:mergers} shows the average number of major ($\mu>0.3$, orange), minor ($\mu=0.1–0.3$, light blue), and marginal ($\mu=0.01–0.1$, gray) mergers in halos per time bin. On average, each dwarf galaxy experiences approximately one major merger and a few minor mergers at $z<7$. In contrast, marginal mergers occur about 10 times more frequently than major and minor mergers combined and therefore account for a nonnegligible fraction of the accreted mass. Consequently, their integrated dynamical influence can be significant, despite their weak individual impact. \citet{HayashiChiba2006} suggested that the total mass of accreted halos primarily determines the thickness of dynamically heated galactic disks. This is consistent with the stronger correlation of $f_\mathrm{acc}$ found for $\mu>0.01$ than for $\mu>0.1$, implying that the high frequency of marginal mergers contributes to this trend. Thus, in dwarf galaxies, not only the accretion of massive halos but also these smaller, more frequent merger events help shape their kinematics.

This finding aligns with recent cosmological zoom-in simulations on ultrafaint dwarf galaxies. \citet{Orkney2021} investigated dark matter core formation in galaxies with halo masses of $M_\mathrm{halo}\sim10^9\,\Msun$ in the EDGE project. They identified two distinct pathways for dark matter heating: stellar feedback at high redshift before reionization, and impulsive heating from minor mergers, which continues even after star formation has ceased. By modifying the initial conditions, they demonstrated that galaxies with late assembly histories experience more late-time minor mergers, leading to continued dark matter heating long after quenching. We observe similar behavior in our sample, albeit at a slightly larger mass scale of $M_\mathrm{halo}\sim10^{10}\,\Msun$.

\subsection{Merger-triggered Gas Disk Formation} \label{sec:gasdisk}

\begin{figure*}[t]
\centering
\plottwo{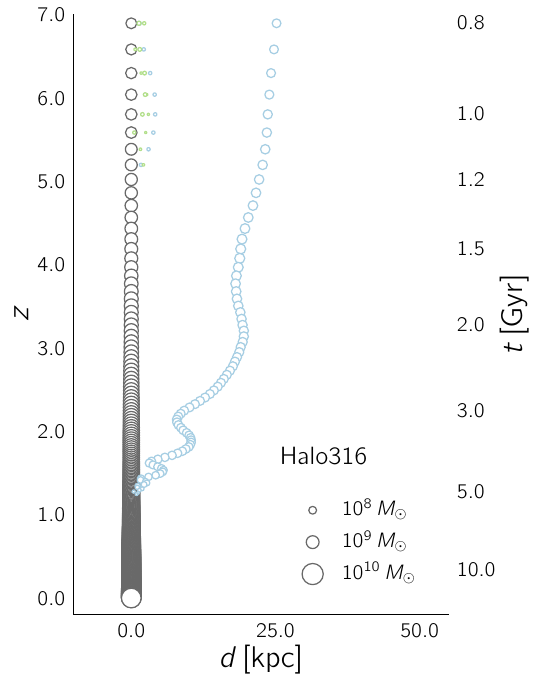}{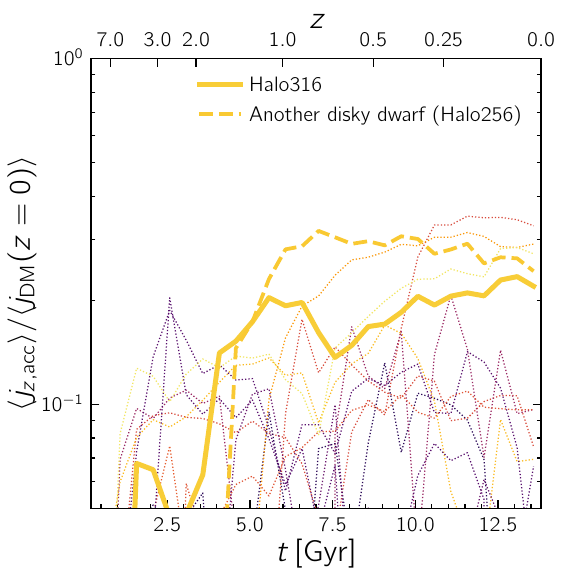}
\caption{\textit{Left panel:} Merger tree of Halo316. The gray circles represent the main halo, while the colored circles indicate the accreted halos ($\mu>0.1$). The $x$-axis shows the distance between the main and accreted halos. Circle size corresponds to subhalo mass. \textit{Right panel:} Gas angular momentum accretion histories for dwarf galaxies in our simulations. The angular momentum is represented as the specific angular momentum aligned with the galaxy's angular momentum vector, normalized by the mean specific angular momentum of the dark matter halo at $z=0$. The lines show the median values within time bins of width $0.5\,\mathrm{Gyr}$. The solid line corresponds to Halo316, which hosts a gas disk. The dashed line represents another galaxy (Halo256), which also forms a gas disk.}
\label{fig:mergertree}
\end{figure*}

To clarify the origin of the strong rotational support of Halo316 ($V_\mathrm{rot,gas}/\sigma_\mathrm{gas}\approx3$) in our simulations, we show its merger tree in the left panel of Figure~\ref{fig:mergertree}. The gray circles represent the main halo, while the colored circles indicate massive accreted halos defined by $\mu>0.1$. Halo316 undergoes a major merger ($\mu\approx0.3$) at $z\approx1.3$. Because this event contributes only modestly to the cumulative accreted mass ($f_\mathrm{acc}\approx0.1$), the associated dynamical heating is unlikely to be significant. Instead, the merger efficiently supplies angular momentum to the main halo and contributes to the formation of a rotationally supported disk.

The right panel of Figure~\ref{fig:mergertree} shows the gas angular momentum accretion histories of dwarf galaxies in our simulations. We plot the component of the specific angular momentum aligned with the galaxy's gas angular momentum vector, normalized by the mean specific angular momentum of the dark matter halo at $z=0$. The quantity is measured for inflowing gas within a radial shell of width $4.5–5.5\,\mathrm{kpc}$ centered on the galaxy.

Halo316 (solid line) exhibits a sharp rise at $z\approx1.6$, corresponding to the last flyby prior to the merger, after which the accreted gas angular momentum becomes well aligned with that of the galaxy by $z=0$. For comparison, another gas-rich disk galaxy in our simulations, Halo256 (dashed line), also undergoes a major merger ($\mu\approx0.4$) at a similar redshift and exhibits a comparable angular momentum accretion history.

This scenario, in which dwarf galaxies form gas disks as a result of late-time major mergers, has also been reported in the FIRE-2 zoom-in simulations \citep{Benavides2025}. In their study, one galaxy develops an ordered gas disk triggered by a major merger with a stellar-to-gas mass ratio of $\sim0.2$ at $z\sim1$.  At $z=0$, it has $M_\mathrm{halo}=4\times10^{10}\,\Msun$ and $M_\star=4\times10^7\,\Msun$.

However, it is important to note that this galaxy is several times more massive than our simulated galaxies. In FIRE-2, no distinct gas disk is observed in galaxies with masses comparable to those in our study ($M_\mathrm{halo}\sim10^{10}\,\Msun$ and $\Mstar=10^6–10^7\,\Msun$) \citep{El-Badry2018}. \citet{El-Badry2018} found that stellar feedback efficiently suppresses the accretion of high-angular-momentum gas at late times and induces large noncircular gas motions in such low-mass dwarf galaxies.

This suggests that the feedback implementation in our simulations may be weaker than in FIRE-2, and star formation may be less bursty. Alternatively, there may be a bias in the sample selection. Our simulations consistently predict stellar masses several times higher than those in FIRE-2 (see Figure~\ref{fig:shmr}). For instance, AGORA-1e10q and AGORA-1e10v are also included in their sample under the names m10q and m10v. According to \citet{Fitts2017}, their stellar masses are $2\times10^6\,\Msun$ and $1\times10^5\,\Msun$, respectively. These values are 3 and 6 times smaller than those predicted by our simulations. Additionally, our star formation threshold $n_\mathrm{H,thres}=1\,\mathrm{cm}^{-3}$ is much smaller than that of FIRE-2 ($n_\mathrm{H,thres}=1000\,\mathrm{cm}^{-3}$ and other thresholds are assigned), so feedback should become weaker. Regarding potential sample bias, their halo concentration parameters, estimated using Einasto profile fitting, appear higher than those in our sample. This suggests that their halos are biased toward more early-assembly systems and therefore might not include those undergoing major mergers at later times.

\section{Conclusions} \label{sec:conclusions}

We have presented results from CROCODILE-DWARF, a suite of cosmological zoom-in simulations of isolated field dwarf galaxies performed with the \textsc{gadget4-osaka} code. By systematically varying the assembly histories of halos with virial masses $M_{200}\sim10^{10}\,\Msun$ at $z=0$, we examined how their formation pathways shape the observed diversity in the kinematic properties of dwarf galaxies within the $\Lambda$CDM framework.

Our simulations successfully reproduce key observed scaling relations, such as stellar-to-halo mass, mass--metallicity, and size--mass, over the stellar mass range $\Mstar=5\times10^6–4\times10^7\,\Msun$. We emphasize that achieving this agreement required a doubled SN energy budget ($\zeta_\mathrm{SN}=2.3\times10^{49}\,\mathrm{erg}\,\Msun^{-1}$) compared to the fiducial \textsc{CELib} output, with energy deposited in both momentum and thermal forms, the latter injected stochastically. This enhancement likely compensates for unresolved early stellar feedback processes and limited numerical resolution.

We find that assembly history plays an important role in determining the structural and kinematic characteristics of dwarf galaxies. Early-forming, high-concentration halos experience rapid star formation and become gas-poor, while late-forming, low-concentration halos maintain high gas fractions ($f_\mathrm{gas}>0.8$) due to delayed star formation and late-time gas accretion. The merger activity experienced by these systems further shapes their gas kinematics, with the trend between gas rotational support and cumulative merger mass fraction suggesting that merger-driven dynamical heating contributes to determining their kinematic state, though the correlation is not statistically significant given our limited sample size of 14 simulations. In rare cases, late-time major mergers promote the formation of extended, ordered gas disks through angular momentum transfer.

Future extensions of this work will first focus on expanding the simulation sample to enable more statistically robust conclusions. A larger suite of simulations, combined with more careful subsample selection, will allow us to isolate the effects of individual assembly parameters on dwarf galaxy properties. We will also explore the role of additional physical processes, including cosmic rays, magnetic fields, and alternative dark matter models such as self-interacting dark matter, on dwarf galaxy evolution, morphology, and kinematics. Combining these simulations with synthetic observational analyses will enable more direct comparisons with upcoming facilities such as \textit{Subaru PFS} and \textit{ARRAKHIS}, as well as future JWST observations of faint galaxies. These studies will further clarify how baryonic physics and assembly history jointly shape the formation and survival of the smallest galactic systems in the Universe.

\begin{acknowledgments}

We thank Keita Fukushima, Nicolas Ledos, and Abednego Wiliardy for many helpful comments. We are also grateful to Volker Springel and others for providing the original version of the \textsc{gadget-4} code, on which the \textsc{gadget4-osaka} code is based. This work used computational resources provided by the SQUID at the D3 Center of the University of Osaka, through the HPCI System Research Project (Project IDs: hp230089, hp240141, hp250119). This work is supported by the JST SPRING, grant No. JPMJSP2138 (K.T.), and MEXT/JSPS KAKENHI grant Nos. JP20H00180, JP22K21349, JP24H00002, JP24H00241, JP25K01032 (K.N.), and JP25K17438 (D.T.). K.N. also acknowledges support from the Kavli IPMU, the World Premier Research Center Initiative (WPI), UTIAS, the University of Tokyo. For the visualization of gas projections shown in Figure~\ref{fig:maps}, we used \texttt{yt} \citep{Turk2011}.

\end{acknowledgments}

\begin{contribution}

K.T. performed all simulations presented in this paper, analyzed the outputs, created the figures, and wrote the first draft. Y.O. has developed the \textsc{gadget4-osaka} code. D.T. and K.N. helped to interpret the results and polished the text together.  

\end{contribution}

\appendix

\restartappendixnumbering
\section{Parameter tuning of SN feedback energy} 
\label{sec:tuning}

\begin{figure}[t]
\centering
\plotone{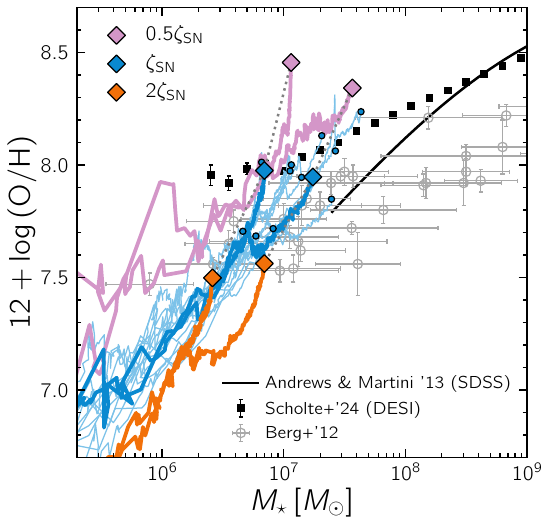}
\caption{Mass--metallicity relation for gas under different SN feedback energies. Results are shown for Halo256 and Halo324.}
\label{fig:tuning}
\end{figure}

Figure~\ref{fig:tuning} shows the mass--metallicity relation for gas at $z=0$, along with the evolutionary tracks, in two samples (Halo256 and Halo324) under different SN feedback energies. We present cases in which the SN feedback energy is half ($0.5\zeta_\mathrm{SN}$, purple diamond) or twice ($2\zeta_\mathrm{SN}$, red diamond) the value used in the simulations (blue diamond), where $\zeta_\mathrm{SN}=2.3\times10^{49}\,\mathrm{erg}\,\Msun^{-1}$. Gray dotted lines connect the same galaxies. We also show the results for other galaxies with the standard SN energy (small blue circles).

We find that the lower SN feedback energy ($0.5\zeta_\mathrm{SN}$) produces metallicities that lie above the observed mass--metallicity relation at $z=0$, indicating insufficient metal ejection from the galaxies. Conversely, the higher-energy case ($2\zeta_\mathrm{SN}$) yields lower metallicities than those of the DESI samples \citep{Scholte2024}. This comparison demonstrates that our adopted SN feedback energy ($\zeta_\mathrm{SN}$) best reproduces the observed mass--metallicity relation at $z=0$.

The SN feedback energy we adopt is twice the standard output of \textsc{CELib}, assuming the Chabrier IMF, but this value remains broadly within the parameter space typically explored in current galaxy simulations \citep{KellerKruijssen2022}. For example, in cosmological zoom simulations, \citet{Applebaum2021} employ a feedback energy of $1.5\times10^{51}\,\mathrm{erg}\,\mathrm{SN}^{-1}$ (corresponding to approximately $1.9\times10^{49}\,\mathrm{erg}\,\Msun^{-1}$) assuming the Kroupa IMF \citep{Kroupa2001}, which exceeds the canonical value of $10^{51}\,\mathrm{erg}\,\mathrm{SN}^{-1}$. Large-volume cosmological simulations such as EAGLE \citep{Crain2015,Schaye2015} and IllustrisTNG \citep{Pillepich2018} implement metallicity-dependent stellar feedback schemes that yield stronger feedback in low-metallicity environments, thereby enhancing feedback in dwarf galaxies.

\bibliography{main}{}
\bibliographystyle{aasjournalv7}

\end{document}